\documentclass[review]{elsarticle}

\makeatletter
\def\ps@pprintTitle{
	\let\@oddhead\@empty
	\let\@evenhead\@empty
	\def\@oddfoot{\reset@font\hfil\thepage\hfil}
	\let\@evenfoot\@oddfoot
}
\makeatother

 \biboptions{comma,sort&compress}

\usepackage{graphicx}
\usepackage{amsmath}
\usepackage{here}
\usepackage{cuted}

\usepackage[dvips]{epsfig}

\begin{document}
\begin{frontmatter}

\title{Colour Confinement and gauge-invariant field-strength correlations.
\,\tnoteref{invit}}
\tnotetext[invit]{Talk given at QCD25 - 40th anniversary of the QCD-Montpellier Conference. }
\author{Adriano Di Giacomo
}
\address{Dipartimento di Fisica Universita' di Pisa and I.N.F.N. Sezione di Pisa\\
3 Largo B. Pontecorvo 56127 Pisa Italy}

\ead{adriano.digiacomo@df.unipi.it}


\date{\today}
\begin{abstract}
 In this paper we produce  evidence that confinement of colour is due to dual superconductivity of $QCD$ vacuum. To do that we put together results of old numerical simulations and results of more recent investigations. The starting point is the  expectation that gauge theories admit a dual description in terms of monopoles. The strategy is then to construct the creation operator  $\mu$ of a monopole and to compute its vacuum expectation value $\langle \mu \rangle$ , the disorder parameter which indicates dual superconductivity.  The Imechanism of   confinement is dual superconductivity of vacuum if  $\langle \mu \rangle \neq 0 $  in the confined phase , and $\langle \mu  \rangle =0$ in the deconfined phase. Confinement  has to be certified by independent methods.

It is shown that gauge invariance  requires that field strengths  be replaced   by gauge invariant field strengths, which are their parallel transports to infinity.
The resulting disorder parameter 
is  a sum of correlation functions of gauge invariant field strength, and its behaviour understood by use of existing lattice data of two-point gauge invariant correlations. 

As a byproduct an apparent existing inconsistency,  the lack of preferred orientation in colour space of the chromo-electric field inside confining flux tubes,
is resolved.

\vspace{.3cm}

\end{abstract}
\end{frontmatter}
\newpage
\section{Introduction}
The Standard Model of elementary particles is an \hspace{.1cm} $SU(3) \times SU(2) \times U(1)$\hspace{.2cm} gauge theory.
 The strong interactions are  described by the subgroup \hspace{.1cm} $SU(3)$\hspace{.1cm}  with quarks in the fundamental representation.  
 The corresponding part of the Lagrangian is known as Quantum Chromodynamics ($QCD$). 
 
 In the usual notations 
\begin{equation}
L_{QCD}=-\frac{1}{4} \big( \frac{1}{N}Tr [G _{\mu \nu} G _{\mu \nu} ] \big)  + \Sigma _f \bar \psi _f( i  D_{\mu} \gamma_{\mu} -m_f) \psi_f  \label{QCD}
\end{equation}
Here $G_{\mu \nu} \equiv \partial _{\mu} A_ {\nu} - \partial _{\nu} A_ {\mu} +i g [A _{\mu}, A_{\nu}]$ is the field strength tensor, $A_{\mu}$ the gauge field, $D_{\mu} \equiv \partial_{\mu} - i g A_{\mu}$ the covariant derivative. $A_{\mu}$ is a matrix in the fundamental representation of the gauge group which we shall assume to be $SU(N)$,  in particular $SU(3)$, $\psi _f$ is the field of the quark of flavour $f$ and mass $m_f$, g the strong coupling constant.

The field theory Eq(\ref{QCD}) is asymptotically free. The effective coupling constant goes small at short distances, and this allows the use of the known methods of perturbation theory. A great amount of data, e.g. of deep inelastic scattering of leptons on hadrons, and of $e^+ e^-$ annihilation into hadrons at high energy have 
been analysed by use of the theory Eq(\ref{QCD}) confirming its predictions  with a good precision,  and in particular establishing beyond any reasonable doubt that quarks do exist.

At large distances the coupling constant becomes large, and the theory unmanageable by analytical methods.  Experimentally an a priori unexpected phenomenon shows up. No free quarks are observed neither in nature nor as a product of  high energy reactions. Empirical upper limits for these observations are very stringent. Typically one has for the number of quarks  $n_q$, \hspace{.2cm}$ n_q \le 10^{-15} n_0$ \hspace{.2 cm}, where $n_0$ is the naive estimate in the assumption that quarks are not confined \cite{dig1}.  

$10^{-15}$ is a small number and the "natural" conclusion is that  $n_q = 0$,  protected by some symmetry or that quarks do not exist as free particles.  Such a phenomenon is known as  CONFINEMENT  OF COLOUR, or simply CONFINEMENT.  Transition from deconfined to confined is a change of symmetry and is expected to be a phaase transition.
\vspace{.2cm}

The basic question is  : HAS  $QCD$ CONFINEMENT BUILT - IN ? 

\vspace{.2cm}

In this paper we  show that the answer to this question is positive. 

The key  property  we will use   to demonstrate  that  is duality \cite{'tH11}.
Duality is a property of field theories and statistical systems described by fields $\Phi$, which have excitations  $\mu$ made stable by topology.
These systems admit two equivalent descriptions:

1) A direct description in terms of the fields $\Phi$ , with coupling constants $g$. $\langle \Phi \rangle$ are called order parameters.

2) A dual description in which the topological excitations $\mu$ become local fields. The coupling constant is $g' \approx \frac{1}{g}$ and $\langle \mu \rangle $ are called disorder parameters.

The prototype system, which is at the origin of this terminology, is the $2d$ Ising model, defined on a two dimensional lattice   in terms of a field $\Phi( \vec i)$ which can assume values $\Phi(\vec i) = \pm 1$, and has nearest neighbour interaction $H = - K \Sigma_{\vec i \hspace{.1cm} \vec j} \hspace{.1cm} \Phi(\vec i) \Phi (\vec j)$ with $K $  a constant  \cite{KW}. In this model the dual topological excitations are the kinks  $ \mu ^{\bar k}$, one-dimensional  (spatial)  configurations $\mu ^{\bar k}( i )= -1, ( i< \bar k )  ,\mu ^{\bar k}(i)= +1, (i \ge \bar k ) $ \hspace{.2cm} and the anti-kinks $- \mu ^{\bar k}$ The dual description is again a $2d$ Ising model  with coupling constant $g'$ a known function of $g$ with $g' \approx  \frac{1}{g} $  at large values of $g $ or  $ g'$ \cite{KW} \cite{KC}.

\vspace{.2cm}

In gauge theories there exist configurations $\mu$  made stable by topology. Their nature depends on the number of spatial dimensions  $d$. For $d=2$ $\mu$ they are vortices, corresponding to a mapping of the circle at infinity on $U(1)$. For $d=3$, which is the physical case, the excitations $\mu$ are monopoles, corresponding to mappings of the sphere at infinity on the group $SU(2)$.  

Monopole configurations are explicitly known. The Dirac monopole for $U(1)$ gauge group \cite{Dirac}.  For higher groups a monopole for each $SU(2)$ subgroup of the gauge group  \cite{'tH1} \cite{Pol}: each of these choices is called  an Abelian Projection.

\section{Confinement by dual superconductivity of the vacuum.}
The idea behind  this mechanism  of confinement, first proposed in \cite{'tH} \cite{m}, is that the dual fields of $QCD$, which are magnetic monopoles, condense in the ground state in the confined phase, in the same way as electrically charged  Cooper pairs condense in the ground state of an ordinary superconductor. In superconductors (at least of type 2)  magnetic field is channeled into  Abrikosov flux tubes with energy proportional to the length, so that magnetic charges are confined. In dual superconductors, where magnetic is interchanged with electric,  the electric charges, the quarks,  should be confined. 

We are not able  to  demonstrate  analytically that this is the case, as already anticipated in Section 1, but we can do it  with the help  of numerical simulations of the theory on lattice. 

The strategy is the following. 

  In the language of duality $QCD$  defined by Eq(\ref{QCD}) is the direct description of the system.  The disorder parameter $\langle \mu \rangle$ can be defined  as the vacuum expectation value of the creation operator of a monopole  in the direct description  and computed by lattice simulations \cite{dig}. At the same time it is possible to detect by independent methods when the system is in the confined phase or in the deconfined one, e.g. by measuring the Wilson loop \cite{Creutz}.
  
   $\langle \mu \rangle \neq 0$  means that the ground state is not an eigenstate of the number of monopoles, i.e. that there is condensation of monopoles or dual superconductivity. One has then to verify that $ \langle \mu \rangle \neq 0$ in the confined  phase , $\langle \mu \rangle =0 $ in the deconfined phase, of course in the thermodynamical limit $V \to \infty$.

The creation operator of a monopole in the direct description of $QCD$  is 
\begin{equation}
 \mu(\vec x , t) = exp  \big ( \int d^3 y    \frac{1}{N} Tr[G_{0 i}(\vec y ,t)T_3] \frac{1}{g} A_{\perp}^i (\vec y -\vec x) \big )  \label{mon} 
\end{equation}

In Eq(\ref{mon}) $\frac{1}{g} A_{\perp}^i (\vec y -\vec x)$ is the vector potential  produced in the location $\vec y$ by a monopole sitting at $\vec x$ in the transverse gauge $\partial _i A^i _{\perp} =0$ .  $\frac{1}{g} $ , the inverse of the coupling constant, is its magnetic charge.  $G_{0i} (\vec x, t) $ is  the $i-th$  space component of the chromo-electric field strength.   $Tr[G_{0 i}(\vec y ,t)T_3] $is its projection on the colour direction, which we denote conventionally by the index $3$, in which the monopole is created. (Abelian Projection).

 In the convolution at the exponent in Eq(\ref{mon}) only the transverse part of $G_{0i}(\vec y ,t)$ survives,  more precisely  of its component in the colour direction identified by the appropriate abelian projection , $3$,   which is the canonically conjugate momentum  to the component of the transverse  field $A_{{\perp}  i}^3  $ in the same colour direction, in whatever quantisation scheme. The operator $\mu$ defined  by Eq(\ref{mon}) shifts any field configuration  $A_{\perp i} ^3 $ by the field  of a monopole $\frac{1}{g} A_{\perp}^i (\vec y -\vec x)$,  in the same way as for a one-dimensional quantum system
 \begin{equation}
 \exp(ipa) |x \rangle = |x +a \rangle
 \end{equation}
 
  i.e. it creates a monopole in the appropriate abelian projection.
  
  The disorder parameter $\langle \mu( \vec x, t)\rangle $   measured on the lattice is given by 
  
  \begin{equation} 
  \langle \mu \rangle = \frac{\int \mu(\vec x ,t) \exp (- \beta S )}{\int \exp( -\beta S)} \label{avmu}
  \end{equation}
 $\beta \equiv \frac{2N}{g^2} $ for gauge group $SU(N)$. The action on the lattice is
 \begin{equation}
 S = \Sigma _{ n , \mu \neq \nu} \Re [ P_{\mu \nu} ( n) -1]  \label{act}
 \end{equation}
 The sum on  $ n$ runs on the $4d$ space-time (the lattice) , and $\mu \neq  \nu$ on the four orthogonal axes of space time.  $P_{\mu \nu}(n)$ is the parallel transport around the elementary square of the lattice sitting at the site $n$ in the hyperplane $\mu \nu$ (Plaquette). In the usual notation
 \begin{equation}
 P_{\mu \nu}( n) = \frac{1}{N}Tr[ U_{\mu} ( n) U_{\nu} ( n+ \hat {\mu}) {U^{\dagger}} _{\mu} ( n+ \hat {\nu}) U^{\dagger} _{\nu} ( n)] \label{plaq}
 \end{equation}
 $U_{\mu} ( n) $ is the parallel transport from the lattice site $\vec n$ to  the neighbour site along the elementary link in the direction $\mu$,  $n+\hat {\mu}$ .
 The measure of the integrals appearing in Eq(\ref{avmu}) is  $\Pi _{n, \mu } dU_{\mu} ( n) $ . The integrals exist and are finite if the group is compact.

 An important thing to notice is  the form of the  monopole operator $\mu$  Eq(\ref{mon}). There is a factor   \hspace{.1cm}$\frac{1}{g}$  \hspace{.1cm}in the exponent coming from the fact that the magnetic charge is proportional  to the inverse of the electric charge, and another factor \hspace{.1cm} $\frac{1}{g}$\hspace{.1cm}  coming from the fact that  the canonical field strength $G_{i 0}(\vec n)$  is  $\frac{1}{g}$ times the lattice field strength. As a consequence the monopole operator has the form  $\mu = \ exp (- \beta \Delta S) $  and
 \begin{equation}
 \langle \mu \rangle =\frac{\int \exp(-\beta (S+\Delta S)}{\int \exp( -\beta S)} \label{mon1}
 \end{equation}
  This holds independent of the gauge group.
  
 It proves convenient to compute, instead of  $\langle \mu \rangle $ itself  the quantity $\rho $  defined as \cite{dp}
 \begin{equation}
 \rho \equiv \frac {\partial log (\langle \mu  \rangle }{\partial \beta} =\langle S \rangle _S  -  \langle (S+ \Delta S) \rangle _{(S+ \Delta S)} \label{rho}
 \end{equation}
 
 The average of the action is very easy to compute numerically. Moreover, since $\langle \mu \rangle =1$ for $\beta =0$
 
 \begin{equation}
 \langle \mu \rangle =\exp \big(\int_0 ^{\beta} \rho(\beta ') d \beta ' \big ) \ \label{murho}
 \end{equation}
 The program  is to measure numerically   $\rho$ and with it the disorder parameter $\langle \mu \rangle$ and to show that it is non zero in the confined phase, zero in the deconfined phase. To do that one has to know the monopole configuration and the corresponding $\Delta S$ of Eq(\ref{mon1}), necessary to  compute  $\rho$ either 
 numerically by use of Eq(\ref{rho}) or analytically by use of the following expansion, proved in the appendix of  \cite{digiacomo1}
 \begin{equation}
  \rho = -\Sigma ^{\infty} _0 \frac{(-\beta)^n}{n!} \langle \langle  (\Delta S)^{n +1}\rangle \rangle   -\langle \langle S \Sigma^{\infty}_1 \frac{(-\beta)^n}{n!}(\Delta S)^n \rangle \rangle \label{rhoan}
 \end{equation}
 In Eq(\ref{rhoan}) the double brackets denote connected correlations.
 
 The disorder parameter $\langle \mu(\vec x, t) \rangle $ is independent on the location $\vec x$ and on the time $t$ at which the monopole is created, due to invariance of the action under space and time translations.
To simplify the notation we shall choose $\vec x =0$ and $t=0$ and denote the disorder parameter simply by $\langle \mu \rangle $ . Moreover $\Delta S$ is non zero by definition  only at $t=0$. We can put \cite{dp} \cite{dlmp}, \cite{digiacomo1}
in analogy with Eq(\ref{act}),
\begin{equation}
S+\Delta S =  \Sigma _{\vec n , \mu \neq \nu} \Re [ P'_{\mu \nu} ( n) -1]
\end{equation}
where \cite{dp} 
\begin{eqnarray}
P'_{\mu \nu} (\vec n, t) &=& P_{\mu \nu} (\vec n, t) \hspace{1.cm} t \neq 0\\
P'_{m n} (\vec n, 0) &= &P_{m n}(\vec n, 0) \hspace{1.cm} m,n =1,2,3.\\
P'_{ i 0} (\vec n, 0) &= & \frac{1}{N} Tr [U_i (\vec n, 0) U_0 (\vec n + \hat i ,0) M_i (\vec n) U_i ^{\dagger} (\vec n, 0 +\hat 0) U_0 ^{\dagger} ( \vec n,0)] \label{pp}
\end{eqnarray}
with 
\begin{equation}
M_i (\vec n) = \exp(\frac{i}{g} A^i_{\perp}(\vec n) T_3)   \label{M}
\end{equation}
Here  $T_3$ is the 3 generator of the $SU(2)$ subgroup in which the monopole lives (Abelian Projection)

By an iterated change of variables in the Feynman integral, of the form $U_i (\vec n , n_0) \to U'_i (\vec n , n_0) = U_i (\vec n , n_0) {M _i }^{-1}(\vec n)$ it is possible to show that  $S+ \Delta S$ describes a system with a monopole added at all the times $t> 0$ \cite{dp} \cite{dlmp}. 

 To  simplify the notation we put $ M_i (\vec n) = C_i (\vec n) + i  S_i (\vec n) T_3$  with
 \begin{eqnarray} 
C_i (\vec n) \equiv \cos ( \frac{1}{g}A^i_{\perp}(\vec n)  \label{ Ci} )\\
 S_i (\vec n) \equiv \sin (\frac{1}{g}A^i_{\perp}(\vec n) ) \label{Si}
\end{eqnarray} 
 This  is true for $SU(2)$ gauge group. 
The extension of the treatment to generic $SU(N)$ group is tedious but trivial.

 We get  from Eq(\ref{pp})
 \begin{equation}
 \Delta S = \Sigma _{\vec n} \Sigma _i  [ C_i (\vec n) -1] \Re P_{i 0}(\vec n) - S_i (\vec n) \hspace{.1cm}\Im Q_{i0} (\vec n) \label{delta}
 \end{equation}
 where
 \begin{equation}
 Q_{i0}(\vec n, 0) \equiv \frac{1}{N} Tr [U_i (\vec n, 0) U_0 (\vec n + \hat i ,0) T_3 U_i ^{\dagger}(\vec n,1) U_0 ^{\dagger} (\vec n , 0)] \label{Qi0}
 \end{equation}
  The formalism can be extended  to the $U(1)$ gauge theory by putting $T_3 =1$ which makes  $Q_{i0}=P_{i0}$. The links in this case are phases.

  For gauge group $U(1)$ the monopole is uniquely identified as the Dirac monopole For higher gauge groups $SU(N )$  it is  highly not trivial  to identify   among infinite possible choices (Abelian Projections) the monopoles which  can condense.
  
\section{ Computing $\rho$ }
 
The quantity $\rho$  Eq(\ref{rhoan}) is a sum of integrals  of correlation functions of fields    multiplied  by numerical factors, extended  to the three dimensional space. $\Delta S$ is non zero only at $t=0$.

The fields are $\Re P_{i0}(\vec n) \approx  G^2_{i 0} (\vec n)$ which is multiplied by the factor $(1-C_i (\vec n))$ and  $\Im Q_{i0} (\vec n) \approx G_{i 0}(\vec n)$
multiplied by the factor $S_i (\vec n)$ . For reasons which will be clear in the following we have exposed the most infrared terms in the expansions of lattice operators in series  of continuum fields. We have also omitted numerical additive terms which do not contribute to connected correlations .

  Terms with odd powers of $S_i(\vec n _i)$ vanish because $Q_{i0}(\vec n, 0)$ is odd under  charge conjugation, which is a symmetry of the system.

 By invariance under translations the correlations of the fields  only depend on the difference of the coordinates. However the arguments $\vec n_k$ of the functions $C_i(\vec n_k) $, $S_i(\vec n _k)$  also depend (linearly)  on their sum $\vec n =\Sigma _i \vec n_i .$  The monopole at the origin breaks translation invariance. 
 
 At large values of $n = | \vec n| $   $A^i_{\perp}(\vec n) \propto \frac{1}{n}$, and as a consequence 
 \begin{eqnarray}
 (1 - C_i (\vec n) )\approx _{n \to \infty} \frac{1}{n^2} \label{Cas}\\
 S_i (\vec n) \approx _{n \to \infty}  \frac{1}{n} \label{Sas}
 \end{eqnarray}
  In Eq(\ref{rhoan}) the integration on $\vec n$ does not involve the correlation functions of  the fields, which only depend on the difference of coordinates, but only the factors $(1-C_i)$ and $S_i$.  Each  $\Delta S$  produces  at large $n$ an even power factor $\propto \frac{1}{n^2}$   due to the term with $(1-C_i )$ and an odd power  factor $\propto \frac{1}{n}$ due to the term with $S_i$, with  the condition that the total number of odd factors has to be even.  As a consequence for  terms of the expansion Eq(\ref{rhoan}) $\propto (\Delta S)^k$ with  $k > 2$ the integration on $\vec n$ is finite and harmless. The first two terms   proportional to $\Delta S$ and $(\Delta S)^2$ instead have a term proportional to $(1- C_i) \propto _{n \to \infty} \frac{1}{n^2}$ and a term proportional to $S_i ^2  \propto _{n \to \infty} \frac{1}{n^2}$ for which  $\int d^3 n $ diverges at large volumes like $V^{\frac{1}{3}}$ independent of  confinement or deconfinement.  If this happens the disorder parameter $\langle \mu \rangle $ does not exist in the thermodynamical limit $V \to \infty $.  We call this kind of infrared divergence a " kinematic divergence".
  
 The only way to get rid of the kinematic divergences, and have a disorder parameter, is that  the coefficient of the diverging part vanishes identically. We shall discuss in detail this point in the next section.
 
 Once we know that kinematic divergences cancel,  the integrals are governed by the behaviour of correlation functions\cite{digiacomo1}. In the confined phase there is a mass gap, integrals are exponentially cut-off at large distances so that $\rho$ is finite and $\langle \mu \rangle \neq 0$. 
 
 Note that  changes $\delta \rho$ of $\rho$ which are such that  $|\int_0^{\beta} \delta \rho (\beta') d\beta' |< \infty $ are irrelevant [Eq(\ref{murho})], because what matters is that the disorder parameter be zero or non-zero, and this property is not changed by such a variation  $\delta \rho$ Eq(\ref{murho}). 
  
 In the deconfined phase there is no scale and everything is governed by dimensional analysis. The correlation function depends on the difference of coordinates.  Each factor  of the form $\int d^3 n_i ( 1- C_i (\vec n_i)) P_{i0}(\vec n _i)$ contributes with dimension in length $(3 -2 -4) = -3 $ or higher negative and the term which contains it can be neglected in the infinite volume limit. Only the terms containing exclusively factors $\int d^3 n_i  {Si} (\vec n _i)Q_{i 0}( \vec n_i)$ which have dimension $(3-2 -1) =0$ can survive and possibly produce a logarithmic divergence in $\rho$ which in turn gives  a power decrease to zero of the disorder parameter $\langle \mu \rangle$ if it is negative. The lowest order term in the expansion Eq(\ref{rhoan}) is $\rho '$\cite{digiacomo1}
 \begin{equation}
 \rho'= \beta \Sigma _{\vec n_1 \vec n_2 i_1 i_2}  \langle \langle \Im Q_{i _1 0} ( \vec n_1) \Im Q_{i_2 0} (\vec n_2)\rangle \rangle S_{i_1}(\vec n_1) S_{i_2 }(\vec n_2) \label{rho'}
 \end{equation}
 By symmetry arguments only terms with $i_1 =i_2$ contribute \cite{digiacomo1}, so that the correlation function is positive definite. 
 To simplify the notation we define 
 \begin{equation} 
 f_i (\vec n_1 - \vec n_2)  =\Sigma _{i} \langle \langle \Im Q_{i  0} ( \vec n_1) \Im Q_{i 0} (\vec n_2)\rangle \rangle
 \end{equation}  
 We add and subtract the term $\Sigma$ which gives the divergence in the integration over $\vec n $ 
 \begin{equation}
 \Sigma =  \beta \Sigma _{\vec n_1 \vec n_2 i }  f_i (\vec n_1 - \vec n_2)S^2_{i}(\vec n)  \end{equation}
 The term with $+$cancels with the other diverging terms and the effective contribution to $\rho$ is 
 \begin{equation}
 \rho' - \Sigma = \beta  \Sigma _{\vec n_1 \vec n_2 i }  f_i (\vec n_1 - \vec n_2)[ S_{i}(\vec n_1) S_{i }(\vec n_2) -S^2_i(\vec n)]  \label{rho'}
 \end{equation}
 $\rho' - \Sigma$ Eq( \ref{rho'}) is negative definite. Indeed $f$ is positive definite and
 \begin{equation} 
 S_{i}(\vec n_1) S_{i }(\vec n_2) \le \frac{1}{2}( S^2_{i}(\vec n_1) +S^2_{i}(\vec n_2) )\le S^2_i(\vec n)
 \end{equation}
 The first of these inequalities is the Schwartz inequality. The second one can be proved as follows . Put   $ \vec \nu \equiv \frac{\vec n_1 - \vec n_2}{2} $  \hspace{.1cm}  \hspace{.1cm} $\vec n_1 = \vec n + \vec \nu , \vec n_2 = \vec n - \vec \nu $.\hspace{.1cm} Since $ (\vec n .\vec \nu ) =0, $  $|\vec n_m | > |\vec n|$ $(m = 1,2)$, and $S_i ^2(\vec n_m) \le S^2 _i (\vec n)$ from Eq(\ref{Sas}).
 
  We can redefine $\rho$ by eliminating all  terms which give a finite contribution to $\int\rho(\beta') d\beta'$ e.g. all the terms coming from correlations including factors of the type $(1- C_i (\vec n))$. We will discuss this point below for $QCD$.
 Note that the derivation is valid for any gauge group.
 
 \section{ $\langle \mu \rangle$ for compact $U(1)$ and for $QCD$}

  In the case of compact $U(1)$ gauge theory we know that the disorder parameter exists in the thermodynamical limit. This is indicated by accurate results on lattice \cite{dp} and by the proof contained in the same paper \cite{dp} that the disorder parameter that  we have defined  is equal to the one defined in the analytic treatment of Ref \cite{FM}, where it was proved  that monopoles do condense, provided there are no electric charges around. Their proof referred to a simplified form of the action, known as "Vilain" action. In ref \cite{pc} the proof was extended to generic lattice action including the Wilson action Eq(\ref{act}), a non trivial statement considering that the theory is a discrete model with no ultraviolet fixed point  which could define a continuum field theory and make terms of higher dimension irrelevant. In Ref \cite{dp} the disorder parameter was measured by use of Eq's (\ref{rho}) and (\ref{murho}) at  different volumes $V$ and found to be non zero and independent of $V$ at large volumes for  $\beta \equiv  \frac{2}{g^2} \le \bar{\beta}$, and to vanish at large volumes as $V^{(-k)}$ for $\beta > \bar{\beta}$ . The value of  $\bar{\beta}$  was found to coincide with that of the deconfining  transition:  the string tension $\sigma$  measured from the large size behaviour of the Wilson loops is non zero for  $\beta \le \bar{\beta}$ and vanishes at
$\beta > \bar{\beta}$\cite{Creutz}. The value of the exponent $k$ is compatible with a first order transition \cite{dp}. All this indicates that the kinematic divergence vanishes in the case of $U(1)$ gauge group.

 This can  indeed be  proved by explicit calculation \cite{digiacomo1} starting from Eq(\ref{rhoan}). The coefficient $D$ of the term proportional to $\int d^3n \frac{1}{n^2}$   which produces the kinematic infrared divergence is given, for generic $SU(N)$ gauge group, by  the expression  $D = \Sigma_k  \frac{(k+1)\beta^{2 k+1}}{(2k)1}D_k$ with 
\begin{equation}
D_k =  \Sigma_i \Sigma _{ \vec n_2} \langle\langle [\Re P_{i 0}(\vec n_1) \Re P_{i 0}(\vec n_2) - \Im Q_{i 0}(\vec n_1) \Im Q_{i 0}(\vec n_2)]S^{2k} \rangle \rangle      \label{Div}
\end{equation}
In the case of $U(1)$ gauge group $Q_{i 0} (\vec n) = P_{ i 0} (\vec n)$. $\Re P_{i 0} = \frac{ P_{i 0} +  P^{*}_{i 0} }{2}$, $\Im P_ {i 0} = \frac{ P_{i 0} -  P^{*}_{i 0} }{2i} $, so that, from Eq(\ref{Div}) 
\begin{equation}
D_k \propto \Sigma_i \Sigma _{ \vec n_2} \langle\langle [ P_{i 0}(\vec n_1)  P_{i 0}(\vec n_2) + P^{*} _{i 0}(\vec n_1) P^{*} _{i 0}(\vec n_2)]S^{2k} \rangle \rangle \label{Div1}     
\end{equation}
 The cross terms in $P_ {i0} P^{*} _{i 0}$ cancel. The two terms in Eq(\ref{Div1})  vanish separately because they carry electric charge   and the charge density vanishes ( $\partial_i F_{i 0} =0$) \cite{digiacomo1} .

 To compute $\rho $ in  the non abelian case ( $SU(N)$ ,$N \ge 2$) one has first to identify the monopoles which condense in the vacuum producing dual superconductivity.  There is  a monopole species for each abelian projection, i.e. for each $SU(2)$ subgroup which can be defined  in the system.  Many choices  have been  proposed during the time e.g.  following the suggestions of Ref \cite{'tH1}.  In lattice simulations the conventional $T_3$ used in updating the configurations was chosen, in the spirit that all abelian projections are equally good.   As a check  criterion the existence of a good thermodynamical limit for the disorder parameter was used: $\langle \mu \rangle $ was expected to show a finite thermodynamical limit at small $\beta$ in the confined phase, and vanish at larger values of $\beta$ in the deconfined phase. 
 
 At the early times lattices were rather small due to  computational limitations and checking thermodynamical  limit was not easy . Numerical results seemed encouraging. Larger lattices\cite{bcdd} and attempts with more exotic groups like e.g. $G_2$ \cite{cddlp} showed difficulties in the limit $V \to \infty$. There was no way to reach a finite thermodynamical limit for $\langle \mu \rangle$ in the confined phase.
 
 Finally the origin  of the problem was understood \cite{digiacomo1}. A theorem proved in \cite{Eli} states that the disorder parameter can not break
 local gauge invariance spontaneously. The operator $\mu$ breaks the gauge symmetry $SU(N)$  to  $SU(N)/U(1)$ if the abelian projection which defines  the monopole is an $SU(2) $ subgroup of the local gauge group. This possibility is forbidden by the theorem in \cite{Eli}. 
 In our treatment this  fact manifests itself by the presence of kinematic infrared divergences, which inhibit the existence of the thermodynamical limit. This is clearly seen in Eq(\ref{Div}) with the definition of $Q_{i0}(\vec x)$ Eq(\ref{Qi0}). The first term   $P_{i 0}(\vec n_1) P_{i 0} (\vec n_2)$ is gauge invariant because the plaquettes are gauge invariant. The second term instead is gauge dependent because $T_3$ is such. The two will never cancel each other. The kinematic divergence does not cancel and no disorder parameter exists. This is a kind of rephrasing  of Elitzur's  theorem in the language of our problem.
 
 The only chance to have a disorder parameter is  to choose an abelian projection such that $T_3$ is gauge invariant \cite{digiacomo1}. That is obtained by parallel transport  of $T_3$ to spatial $ \infty$.  This is a necessary but generally not sufficient condition to cancel kinematic infrared divergences.
 \begin{equation} 
  T_3 \to  V_C (\vec x, \infty) T_3 V^{\dagger} _C (\vec x, \infty)
 \end{equation}
 Looking back at Eq(\ref{mon}) this is equivalent to replace in all what follows the field strength $G_{0 i}(\vec x) $ by its parallel transport  to  $\infty$. This brings to the definition of  the gauge invariant field strengths  $F_{\mu \nu} (\vec x) $  by the formula 
  \begin{equation}
 F_{\mu \nu} (\vec x) =  V _C^{\dagger}  (\vec x , \infty)  G_{\mu \nu} (\vec x) V_C (\vec x, \infty) \label{F}
 \end{equation}
 An  arbitrary  choice of the path $C$  for  each point  $\vec x$   gives a gauge invariant field strength $F_{\mu \nu} (\vec x)$ and a gauge invariant contribution to the divergent part Eq(\ref{Div}). If, however, we require the field  $ F_{\mu \nu} (\vec x) $ to be differentiable in $\vec x$ the choice is strongly restricted: e.g. neighbouring sites must have the same path, except for the small displacement between them. If that happens it is trivial to prove \cite{digiacomo1} that
 \begin{equation}
 \partial_{\rho} F_{\mu \nu}(\vec x)  = V _C^{\dagger}(\vec x , \infty) D_{\rho} G_{\mu \nu} (\vec x) V_C (\vec x, \infty)
 \end{equation}
 In particular, putting $\mu = \rho$ ,and  in absence of quarks ( $D_{\mu} G_{\mu \nu} =0$ ) we have
 \begin{equation}
  \partial _{i} F_{i  0} =0
\end{equation}
In addition, if we require the gauge invariant field strengths  $F_{\mu \nu} $ to be differentiable, it can be proved  by an argument of continuity that any finite number of them say $k$ must have the paths $C$ overlapping  each other after some common point \hspace{.1cm} $\vec O$ \hspace{.1cm} in their way to spatial infinity \cite{digiacomo2}.  The overlapping  part of the paths can be integrated over in the Feynman integral which defines the theory in the strong coupling expansion, producing a sum of products of $\delta _{a b}$ 's and $\epsilon _{a b c}$ 's of the colour indices $a, b, c,...$ of the parallel transport lines from $\vec O$ to the field locations .  That sum does not depend on $\vec O$ nor on the form of the line $C$ but only depends on the gauge group and on the number of fields \cite{Creutz2}\cite{digiacomo2}.
 The correlation function of gauge invariant fields is a definite combination  of  gauge invariant correlation functions of gauge fields  in the language of Stochastic  vacuum model of $QCD$ \cite{Dosch} \cite{Simonov}. The point $\vec O$ plays the role of the arbitrary point in which all the fields are parallel transported to make correlation functions gauge invariant in the stochastic model. It is easy to verify that its position is irrelevant. The parallel transport paths from the fields to it are still arbitrary. 
 If we choose them such that the parallel transport between the fields in the two point correlation function be a straight line we can prove that the kinematic divergences cancel and the disorder parameter $\langle \mu \rangle$ exists . We refer to \cite{digiacomo1} for the proof.  
 
  In the confined phase there is a mass gap and the integrals  on the relative coordinates converge because the correlation functions vanish exponentially at large distances. They can then be put equal to zero because  their  contribution is  like  a finite $\delta \rho$, giving a non zero multiplicative  factor to the disorder parameter .
  
   In the deconfined phase there is no mass scale and the correlation functions are sums of inverse powers of the relative coordinates, with powers dictated by their physical  dimension. The fields  in the correlation of the form $P_{i 0} (\vec n)$ multiplied by the factor $(1 - C_i (\vec n))$ contribute a dimension $-3$  [$d^3n \frac{1}{n^2} G^2 _{i 0}$], The fields of the form $Q_{i 0}(\vec n) Si(\vec n)$  contribute a dimension $0$ [ $d^3 n \frac{1}{n} G_{i 0}]$. Correlation functions containing at least one factor multiplied by $(1- C_i (\vec n)$ give a vanishing contribution to $\rho$ in the infinite volume limit and can be neglected. 

In conclusion only the terms in  the expansion Eq(\ref{rhoan}) of $\rho$ which contain only $Q_{i 0} S_i(\vec n)$  factors are left, which have dimension $0$ and can produce a negative logarithmically divergence at large volumes in the deconfined phase , i.e. a zero value of the disorder parameter $\langle \mu \rangle $. The lowest order relevant term in the expansion is $\rho'$ of Eq(\ref{rho'}). We have shown from general principles in Section 3 that it is negative definite as requested. We have argued  that the absence of kinematic divergences requires that the parallel transport connecting the fields in a two point connected correlation function is along a straight line. Such correlations are well known from numerical simulations \cite{dp} \cite{ddss} since a long time, also at finite temperature \cite{dmp}. They are positive definite, decrease exponentially at large distances and vanish at the deconfining transition \cite{dmp}. This is an important point  observed in numerical simulations \cite{dmp} because it implies that the disorder parameter is non zero in the confined phase and vanishes exactly at the deconfining transition. Apparently the quantity $\rho'$ defines by itself a disorder parameter which demonstrates confinement of colour by dual superconductivity.

This is true if the contribution of higher order terms of dimension zero in the expansion of $\rho$ Eq(\ref{rhoan}) are irrelevant.
We do not know much about such correlations neither from numerical simulations  nor from phenomenology. Neglecting them seems reasonable,  in the line of what is done in the Stochastic vacuum model of $QCD$ \cite{Dosch} \cite{Simonov}.  The problem deserves further study.

\section{Conclusions}
We have reported on the status of the idea that confinement of colour in strong interaction is due to dual superconductivity of the vacuum. The strategy is to construct the creation operator of  monopoles and use its vacuum expectation value as a disorder parameter. 
Everything works at a rigorous level in $U(1)$ gauge theory. Vacuum is a dual superconductor in the sense that the disorder parameter is non zero in the confined phase, zero in the deconfined one.  

In $QCD$ difficulties in the identification of the monopoles  which condense have been  resolved by better understanding the role of gauge invariance. Field strengths get replaced by gauge invariant field strengths.  The price to pay is that the disorder parameter is not directly measurable in numerical simulations, but can only be determined as a series of integrals of gauge invariant correlations of increasing order $n$.  The contribution of most of them   is well understood. In particular a subset of them is identified   which can  produce a zero value of $\langle \mu \rangle $ in the deconfined phase. The lowest order term of this subset, the integral of the 2-point function of gauge invariant chromo-electric fields, is well known from old lattice simulations, also at non zero temperature up to the deconfining phase transition. The contribution to the disorder parameter is non zero up to the deconfining transition, and it is proved  that it vanishes as a power of $V$ in the limit $V \to \infty$ in the deconfined phase.  Higher order terms which could on dimensional basis behave in the same way are not understood neither numerically nor analytically. In the spirit of the stochastic vacuum approach they should be negligible. Further studies will  possibly clarify the issue. 

Finally we recall an apparent difficulty  of the mechanism of confinement by dual superconductivity of the vacuum, studied in the early times in \cite{GW} and in \cite{ddmo}.  In flux tubes connecting  coloured sources the  orientation in colour space of the chromo-electric field should be related to the colour orientation of the abelian projection in which the monopoles live. Even if we do no know that  abelian projection, the colour distribution of the field in the  flux tubes should remember it. The observed distribution instead is compatible with  flat. This fact was considered  a problem for the mechanism of confinement by dual superconductivity \cite{GW} and would be such if the abelian projection were identified by an operator transforming in the local gauge group. Instead the preferred colour direction is a parallel transport of $T_3$ from spatial infinity, and can vary from point to point.

\end{document}